# Atomically flat reconstructed rutile $TiO_2$(001) surfaces for oxide film growth

Y. Wang,[1] S. Lee, [1] P. Vilmercati,[2,3] H. N. Lee, [1] H. H. Weitering,[3,1] and P. C. Snijders[1,3]

[1]Materials Science and Technology Division, Oak Ridge National Laboratory, Oak Ridge, Tennessee 37831, USA

[2]Joint Institute for Advanced Materials at The University of Tennessee, Knoxville, Tennessee 37996, USA

[3]Department of Physics and Astronomy, The University of Tennessee, Knoxville, Tennessee 37996, USA

**Abstract**

The availability of low-index rutile $TiO_2$ single crystal substrates with atomically flat surfaces is essential for enabling epitaxial growth of rutile transition metal oxide films. The high surface energy of the rutile (001) surface often leads to surface faceting, which precludes the sputter and annealing treatment commonly used for the preparation of clean and atomically flat $TiO_2$(110) substrate surfaces. In this work, we reveal that stable and atomically flat rutile $TiO_2$(001) surfaces can be prepared with an atomically ordered reconstructed surface already during a furnace annealing treatment in air. We tentatively ascribe this result to the decrease in surface energy associated with the surface reconstruction, which removes the driving force for faceting. Despite the narrow temperature window where this morphology can initially be formed, we demonstrate that it persists in homoepitaxial growth of $TiO_2$(001) thin films. The stabilization of surface reconstructions that prevent faceting of high-surface-energy crystal faces may offer a promising avenue towards the realization of a wider range of high quality epitaxial transition metal oxide heterostructures.

## Introduction

Most transition metal dioxides ($MO_2$, with M denoting a 3*d*, 4*d*, or 5*d* transition metal ion) have the rutile structure. They exhibit properties that are the subject of intense attention in condensed matter physics.[1-7] Moreover, their technological value for e.g. catalysis,[8-10] gas sensing,[11] and as (transparent) conductors[11-13] cannot be overstated. Tailoring these properties by confining these materials to thin film geometries and e.g. subjecting them to epitaxial strain requires pseudomorphic growth on suitable substrates. Indeed, the preparation of atomically flat $SrTiO_3$ (001), (110), and (111) substrates[14-16] has enabled the growth of a large range of perovskite oxide heterostructures with precisely controlled atomic thicknesses, thereby facilitating the discovery of many exciting properties.[17-19] Similarly, for rutile oxide film growth, the availability of different low-index atomically flat rutile $TiO_2$ substrates is a key requirement for advancing the utility of rutile transition metal oxide heterostructures in condensed matter and applied physics. To date, most rutile oxide heterostructures are grown on $TiO_2$(110) substrates. It is very difficult to prepare $TiO_2$ (100) and (001) surfaces that are crystalline, atomically flat, and stable against faceting, because their surface free energies are high.[20]

Rutile $TiO_2$ has a body-centered tetragonal unit cell ($a$ = $b$ = 0.458 nm, and $c$ = 0.295 nm), Fig. 1(a), with two formula units per unit cell. The Ti atoms are octahedrally coordinated by oxygen atoms. In the [001] direction, i.e. along the rutile the $c$-axis, the crystal structure can be described by alternating charge neutral $TiO_2$ layers, as shown in Fig. 1(b). The layers stack in such a way that alternating atomic layers are rotated 90° with respect to one another. Their separation is 0.147 nm, i.e. $c$/2. Consequently, on an atomically flat bulk-terminated surface, steps with half-unit-cell height are expected to be present. However, the customary substrate cleaning approach of sputtering and annealing in ultra-high vacuum (UHV), widely used for the (100), (110) and (011) orientations,[21-24] introduces either a faceted morphology on $TiO_2$(001), see Fig. 1(d),[25-27] or small (~20 nm) terraces with a ridge-like structure.[28] The resulting surface is unsuitable for epitaxial growth of rutile films with an atomically precise thickness or a well-defined (biaxial) strain state.[29] This instability towards faceting is caused by the high surface energy of the $TiO_2$(001) surface due to the large number of broken bonds on this surface.[20,29,30] While some recent publications[30-32] reported flat surfaces of $TiO_2$(001), no evidence was presented concerning the stability of this morphology under conditions that are typical for oxide film growth. Moreover, there are no reports demonstrating the persistence of an atomically flat 'step-terrace' morphology after homoepitaxial film growth on $TiO_2$(001).

Here we present a method to prepare atomically flat rutile $TiO_2$(001) substrate surfaces with a step-terrace morphology. We find, using LEED, RHEED, and STM, that the surface is terminated by a two-domain $c(7\sqrt{2}\times5\sqrt{2})R45°$ surface reconstruction, featuring stripes along the [110] and [-110] directions. The step-terrace morphology survives annealing in an oxygen plasma or in vacuum at typical oxide growth temperatures. Moreover, we have been able to grow homoepitaxial films on this surface and demonstrate that the atomically flat step-terrace morphology is retained. We conjecture that the observed stability of the step-terrace morphology of the substrates is enabled by a decrease in surface energy due to the formation of the surface reconstruction.

**Experimental methods**

$TiO_2$(001) substrates (CrysTec, miscut angle < 0.5°) with dimensions 5×5×0.5 $mm^3$ were first cleaned by treating them in isopropyl alcohol (IPA) ultra-sonication baths for 3 min to remove organic contaminants. Next, the substrates were rinsed in deionized (DI) water to remove any residual IPA from the surface. The substrates were then etched in buffered hydrofluoric acid ($NH_3F$ : HF = 10:1, JT Baker) for 10 min to aid terrace formation, followed by a DI water rinse. This effect was ascribed to impurity removal from the $TiO_2$ surface.[30] The $TiO_2$ substrates were subsequently annealed in a tube furnace in air or ultra-high purity oxygen. An alumina boat was used to load the substrates. After inspecting the surface morphology in ambient conditions with an atomic force microscope (AFM), the substrates were transferred into a UHV system for LEED, RHEED, oxygen plasma exposure and STM experiments. During oxygen plasma exposure at a chamber pressure of $5 \times 10^{-5}$ mbar and a plasma current of 30 mA, samples were heated by a laser heater. The temperature was monitored with an infrared pyrometer calibrated using a thermocouple on a dummy sample surface. X-ray photoelectron spectroscopy (XPS) was performed in a separate system using monochromatized Al *K*α x-rays. Homoepitaxial growth of $TiO_2$(001) films by pulsed laser epitaxy (PLE) was conducted in a separate system to illustrate the usefulness of the atomically flat substrate for epitaxy, employing a substrate temperature of 300 °C, and an oxygen pressure of $2 \times 10^{-2}$ mbar. Additional experimental details on the PLE growth are available in Ref. 33. All AFM, STM, XPS, LEED, and RHEED data were taken at room temperature.

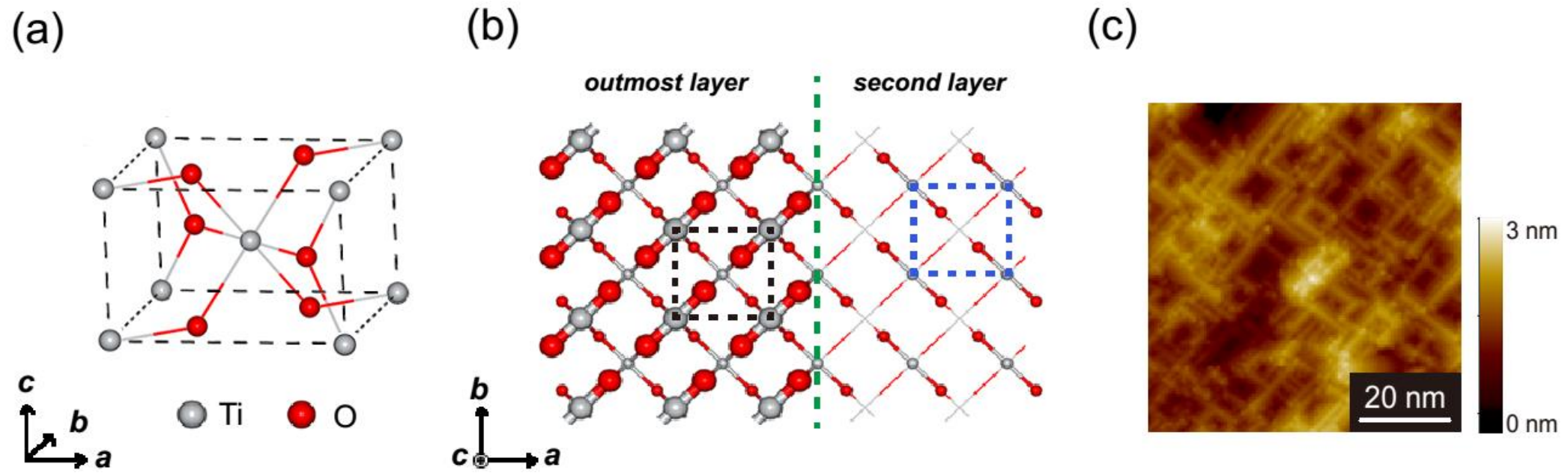

*Figure 1. (a) Ball and stick models of the rutile $TiO_2$ unit cell. (b) Top view of the bulk terminated $TiO_2$(001) surface.. Gray and red balls represent Ti and O atoms, respectively. Larger (smaller) balls are located in the first (second) layer. Dashed rectangles indicate the 1 × 1 bulk terminated unit cells in each layer. (c) STM image of the faceted surface morphology after $Ar^+$ sputtering (2 kV, 20 mA filament emission current, 20 min.) and UHV annealing ( ~ 800 °C, 20 min.).*

**Results and discussion**

In Figure 2 we present the evolution of the surface morphology as measured with AFM for $TiO_2$(001) as a function of annealing temperature in air, systematically covering under-annealed and over-annealed conditions in order to establish clear references for future reproducibility. A temperature ramp rate of 15 °C/min was used, and the duration of the thermal annealing was kept at 5 hours. After wet chemical treatment and before the furnace anneal, the surface is smooth: no step edges are visible and the RMS roughness measured in a 25 $\mu m^2$ area is ~ 0.76 nm (Fig. 2(a)). At an annealing temperature of 780 °C, discontinuous flat terraces emerge (Fig. 2(b)). The area coverage of these terraces increases until the temperature reaches 820 °C where the surface is uniformly covered by atomically flat terraces with an average width of about 400 nm (Fig. 2(d)). At 850 °C, faceted islands appear, with the faceted structures showing as bright elongated features along two perpendicular directions (Fig. 2(e)). Upon increasing the annealing temperature to 900 °C, the surface is fully covered with these faceted islands, consistent with a previous report.[29] Fig. 2(d) and (f) reveal uniform surface morphologies with RMS roughness of 0.18 nm and 1.09 nm (in a 25 $\mu m^2$ area) respectively, showing a large increase in roughness above the optimal annealing temperature. We conclude that wet chemical treatment followed by a five-hour anneal at 820 °C in air produces stepped surface morphologies with atomically flat terraces (Fig. 2(d)) that can serve as an excellent template for subsequent film growth.

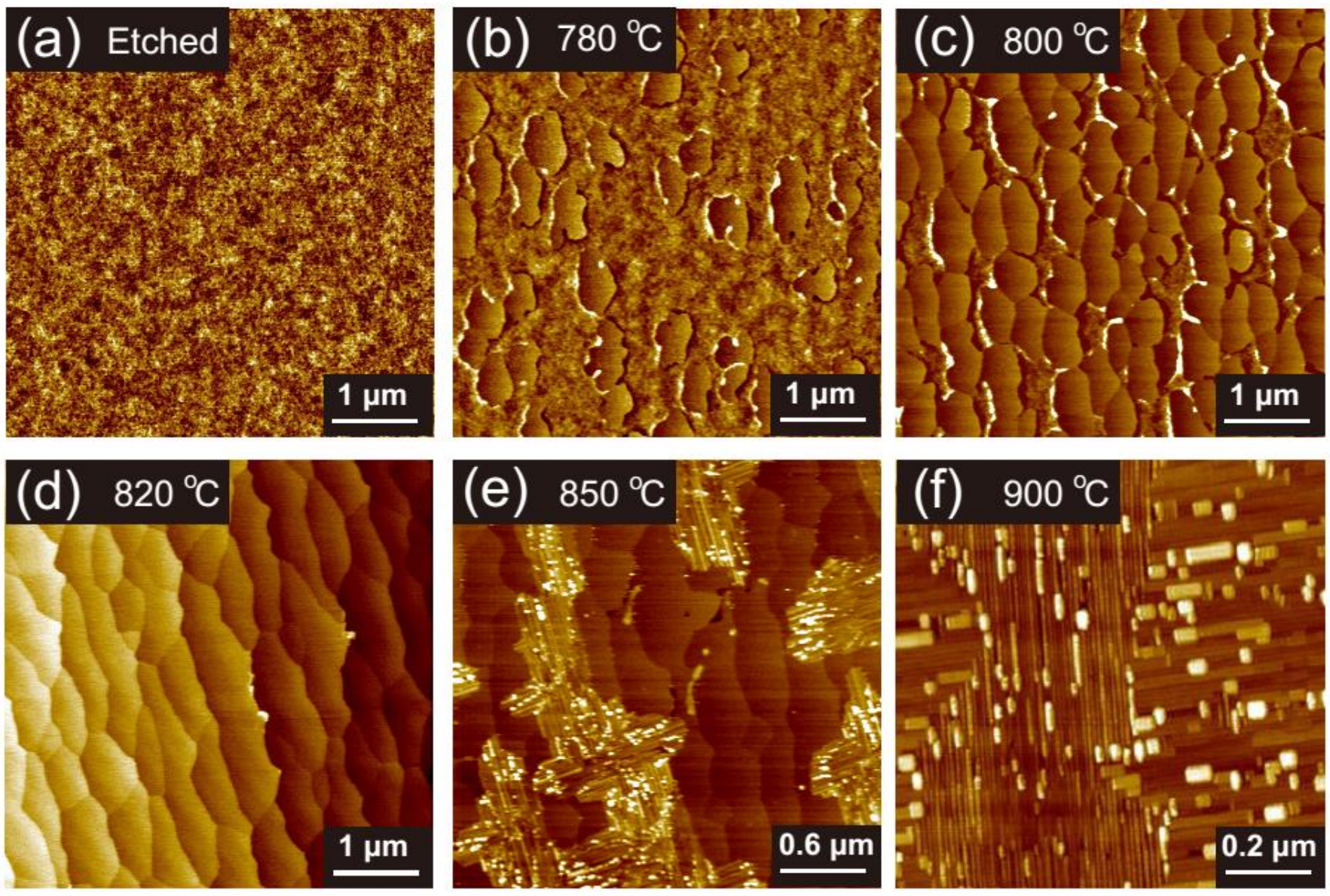

*Figure 2. Ambient AFM images after annealing substrates in air. Starting from a chemically treated $TiO_2$(001) substrate (a), the sample is sequentially annealed in air for 5 hours at (b) 780 °C, (c) 800 °C, (d) 820 °C, (e) 850 °C, and (f) 900 °C.*

These atomically flat $TiO_2$(001) substrates were transferred into a UHV system. STM experiments on these surfaces are difficult due to the insulating nature of the sample, but LEED and RHEED can be performed without any further treatment. Besides the integer order spots, additional fractional order spots are visible in the LEED and RHEED patterns, although they are very faint (see Fig. S1(a)[33]), showing that a surface reconstruction forms already during the annealing treatment in the furnace. This surface reconstruction is further characterized below. To ascertain whether it is contamination from the annealing treatment in air that stabilizes the observed reconstruction and step-terrace morphology, we repeated the annealing treatment using ultra-high purity oxygen gas instead of air. This resulted in LEED and RHEED patterns showing the same superstructure (Figs. S1(b)[33]), indicating that the reconstruction most likely is not induced by e.g. carbon, nitrogen, or other contaminants from the air during the annealing treatment. Indeed, X-ray photoelectron spectra taken from an air-annealed substrate before any degassing in the vacuum system show the presence of only nitrogen and carbon, and only in small quantities that are expected after a transfer through air (Fig. S2a)[33]). This is further corroborated by the absence of a $Ti^{3+}$ component in the Ti 2*p* core level spectrum of the as-loaded sample (Fig. S2(c)[33]) showing that nitrogen doping is indeed absent,[4,33,34] and that the Ti ions are fully oxidized. Finally, a vacuum anneal at 500 °C removes nearly all N and C intensity from the XPS survey spectra (Fig. S2(a)), and does not decrease the quality of the surface diffraction pattern (Fig. S2(b)). These results thus conclusively reveal that the reconstruction already forms during the furnace anneal, that it is a $TiO_2$ structure, and not due to ordered adsorbates.

After exposing the surface to atomic oxygen using an *in situ* oxygen plasma treatment at 500 °C for 40 minutes, the fractional order spots in the electron diffraction patterns become very clear, see Fig. 3(a) and Fig. S3.[33] The superlattice spots can all be indexed using a $c(7\sqrt{2}\times5\sqrt{2})R45°$ reconstruction, with two 90° rotated domains as dictated by the four-fold symmetry of the (001) substrate (Fig. S4[33]). Fig. 3(b) and (c) display (15 kV) RHEED patterns after $O_2$ plasma treatment with the electron beam incident along [010] and [110], respectively. Kikuchi lines and integer order spots in the 0th and 1st Laue zones are visible, some of which are indexed as indicated in the yellow dashed circles. Many superlattice spots due to the $c(7\sqrt{2}\times5\sqrt{2})R45°$ surface structure are observed, consistent with the LEED data.

To improve the electrical conductivity for STM experiments, we mildly degassed as-loaded samples at ~150 °C in vacuum. The STM image in Figure 3(d) shows that the as-loaded surface retains the step-terrace morphology after this degassing (and before the oxygen plasma treatment mentioned above). Line profiles distinguish steps with a height of ~ 0.15 nm (green lines in Fig. 3(d) and (f)) and ~0.3 nm (blue lines in Fig. 3(d) and (f)), which correspond to half and full unit cell height steps, respectively. The presence of half unit cell height steps is consistent with the crystal structure of a (001) surface plane (see Fig. 1). The RMS roughness on the terraces (as measured in an area of 2500 $nm^2$) is ~ 0.08 nm. This shows the terraces are atomically flat even though they do not reveal well-defined atomic order in STM imaging, see also Fig. S5,[33] despite LEED already showing a faint reconstructed pattern, Fig. S1.[33] To observe atomic order in STM imaging, annealing in an $O_2$ plasma is necessary. This also improved the electron diffraction patterns in Figs. 3(a-c), as discussed above. The plasma anneal was followed by a mild vacuum anneal at 150 °C, if necessary, to recreate the higher surface conductivity needed for the STM experiments. The STM image in Fig. 3(e) shows that each terrace is uniformly covered by periodic stripes separated by ~ 1.6 nm, i.e. $\frac{5}{2}\sqrt{2}a$, consistent with the superstructure having a centered unit cell. The atomic scale disorder visible on the surface could be due to surface oxygen vacancies from the vacuum anneal. As expected, the height difference between two terraces is $c/2$ when the stripes on adjacent terraces are rotated by 90°. This reconstruction is observed in STM over a relatively wide range of oxygen plasma annealing temperatures ranging from 380 °C to at least 600 °C, while the surface remains atomically flat. This indicates that this reconstruction is quite stable against faceting, which is highly significant as most commonly grown rutile oxide films have optimal growth temperatures below 600 °C.

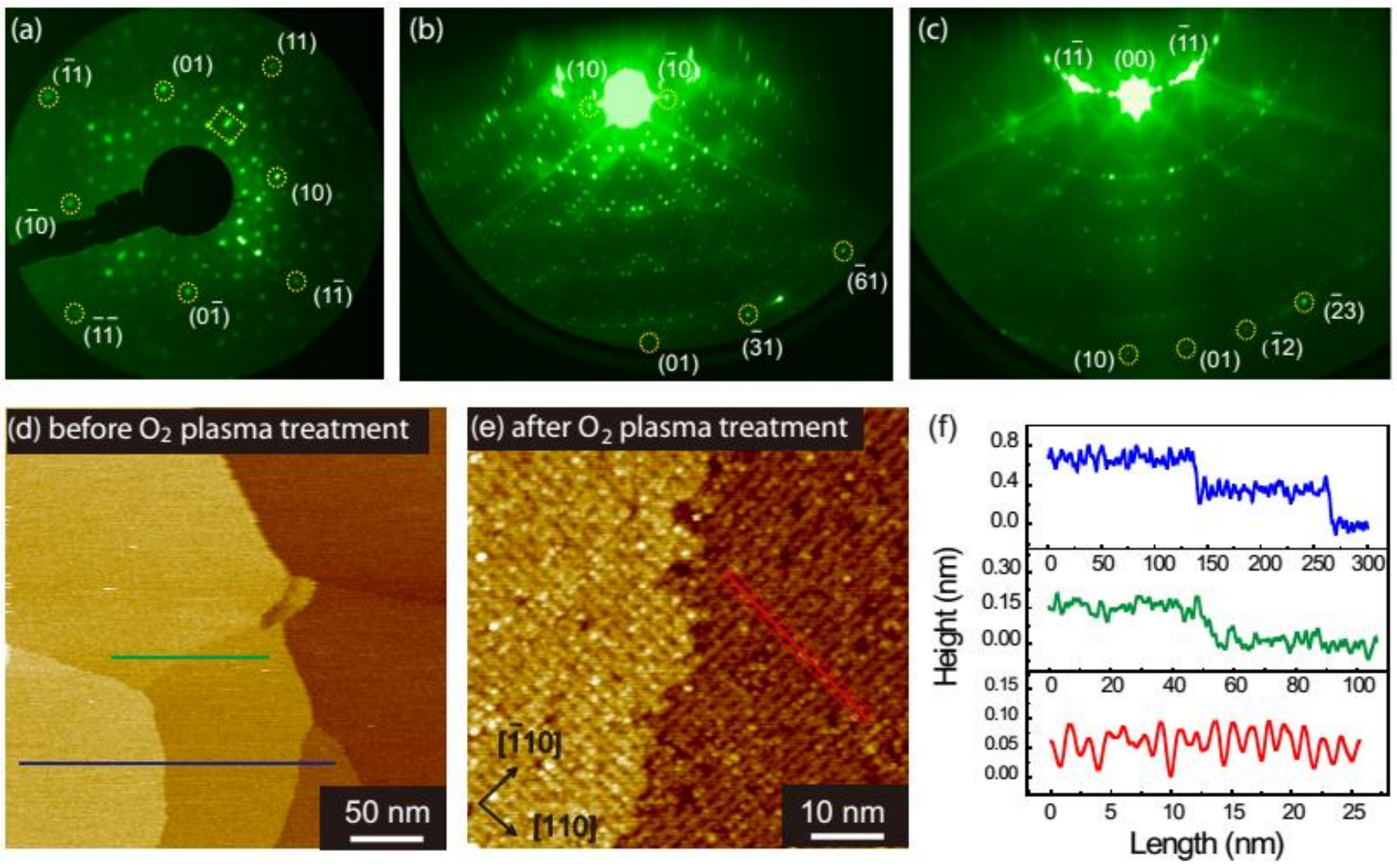

*Figure 3. (a) LEED pattern acquired after oxygen plasma exposure at 500 °C for 40 minutes. Integer order spots are indicated by yellow dashed circles. A superlattice unit cell is indicated by a yellow dashed rectangle. (b) and (c) RHEED patterns taken along the [010] and [110] directions, respectively.*

*Some integer order spots are indicated by yellow dashed circles. Corresponding STM images and height profiles before (+5.0 V, 5.0 pA, (d)) and after (+6.0 V, 5.0 pA, (e))* $O_2$ *plasma treatment. No atomic order is visible on the terraces before* $O_2$ *plasma treatment (d) (see also Fig. S5); it appears in STM imaging only after* $O_2$ *plasma treatment (e). (f) STM height profiles measured in the locations indicated with the green and blue (d), and red (e) lines in the respective STM images. The oscillations in the red line reveal the 1.6 nm periodicity perpendicular to the stripes of the reconstruction in (e).*

Next, we grew ~6 monolayers of $TiO_2$ on the $c(7\sqrt{2}\times5\sqrt{2})R45°$ reconstructed $TiO_2$(001) substrate using PLE. Since film growth was performed in a separate vacuum system, we verified that the reconstruction persists even after (re-)exposure to air, and hence confirmed that the same substrate condition existed for PLE growth (Fig. S6[33]). The morphology of the grown film, imaged with AFM (Fig. 4), shows the same step-terrace morphology after growth. We note that such results, demonstrating the retention of an atomically flat step-terrace morphology upon homoepitaxial overgrowth of a $TiO_2$(001) substrate, have not been reported before. RHEED patterns acquired directly before and after growth (insets in Fig. 4) confirm that there are no significant differences in surface structure and morphology. The AFM and RHEED data clearly illustrate that atomically flat (001)-oriented films can be grown epitaxially on properly prepared $TiO_2$(001) substrates, not hindered by the large unit cell reconstruction of the starting substrate. This demonstration of the possibility for epitaxial (001) rutile film growth while maintaining a well-ordered morphology of atomically flat terraces, has the potential to enable the growth a wider variety of atomically precise rutile heterostructures.

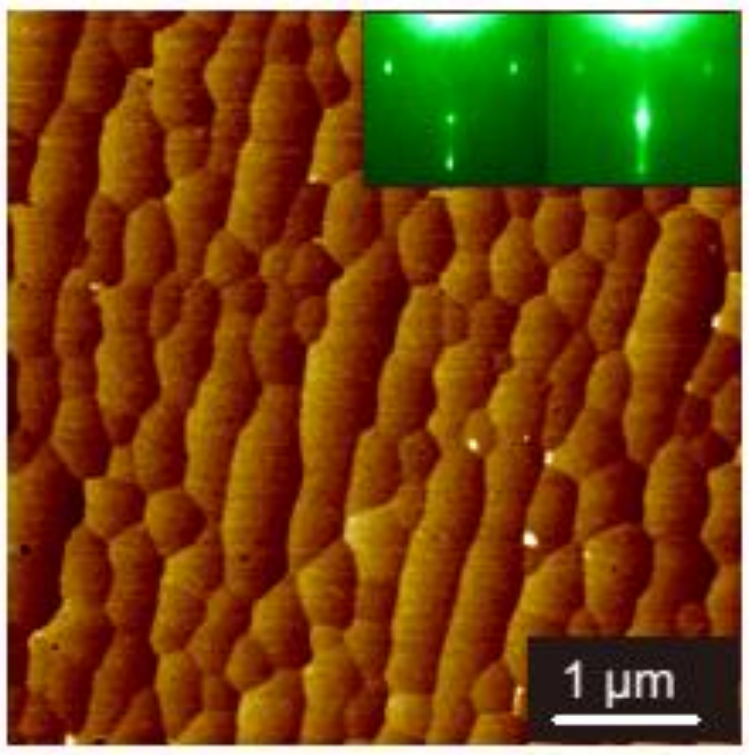

*Figure 4. Ambient AFM image of a ~6 monolayer thick homoepitaxial* $TiO_2$*(001) film grown by PLE. Insets: RHEED patterns recorded before (left) and after (right) homoepitaxial growth.*

In conclusion, we have identified a narrow thermal annealing temperature window to prepare atomically flat and relatively stable $TiO_2$(001) surfaces, in spite of their high surface energy and the associated instability towards faceting. During this annealing the surface reconstructs into a superstructure with a $c(7\sqrt{2}\times5\sqrt{2})R45°$ unit cell that is stable towards faceting up to at least 600 °C. We conjecture that this surface reconstruction decreases the intrinsically high surface energy of the $TiO_2$(001) surface sufficiently to remove the driving force towards faceting. Indeed surface reconstructions have been exploited before to gain access to atomically flat morphologies of crystal faces that are otherwise instable towards a faceted morphology,[35,36] and to prevent roughening or faceting transitions in homo- and heteroepitaxial growth.[37] While these results made use of foreign elements acting as surfactants in vacuum, our atomically flat, reconstructed $TiO_2$(001) surface is obtained in air and without the introduction of surfactant impurities. Finally, we furthermore demonstrated that these atomically flat $TiO_2$(001) substrates are good substrates for growing high quality epitaxial $TiO_2$ films. Our results may pave the way for the realization of a wider range of atomically precise epitaxial transition metal oxide heterostructures in general, and specifically for (001) oriented rutile oxides.

**Acknowledgement**

This work was supported by the U.S. Department of Energy, Office of Science, Basic Energy Sciences, Materials Sciences and Engineering Division. We thank G. Eres for useful discussions.

# Supplementary Material for “Atomically flat reconstructed rutile $TiO_2$(001) surfaces for oxide film growth”

Y. Wang,[1] S. Lee, [1] P. Vilmercati,[2,3] H. N. Lee, [1] H. H. Weitering,[3,1] and P. C. Snijders[1,3]

[1]Materials Science and Technology Division, Oak Ridge National Laboratory, Oak Ridge, Tennessee 37831, USA

[2]Joint Institute for Advanced Materials at The University of Tennessee, Knoxville, Tennessee 37996, USA

[3]Department of Physics and Astronomy, The University of Tennessee, Knoxville, Tennessee 37996, USA

**Pulsed Laser Epitaxy growth parameters**

We ablated a sintered $TiO_2$ target using a KrF excimer laser ($\lambda$ = 248 nm, repetition rate 10 Hz, laser fluence 1 J/cm$^2$). The oxygen pressure and substrate temperature were optimized at 2 × 10$^{-2}$ mbar and 300 °C, respectively. The distance between target and substrate was kept at 45 mm, and the laser spot size on the target was approximately 0.49 mm × 0.80 mm. Film growth started after pre-ablating the target for 1000 pulses. The deposition rate was determined as 111 pulses/ML.

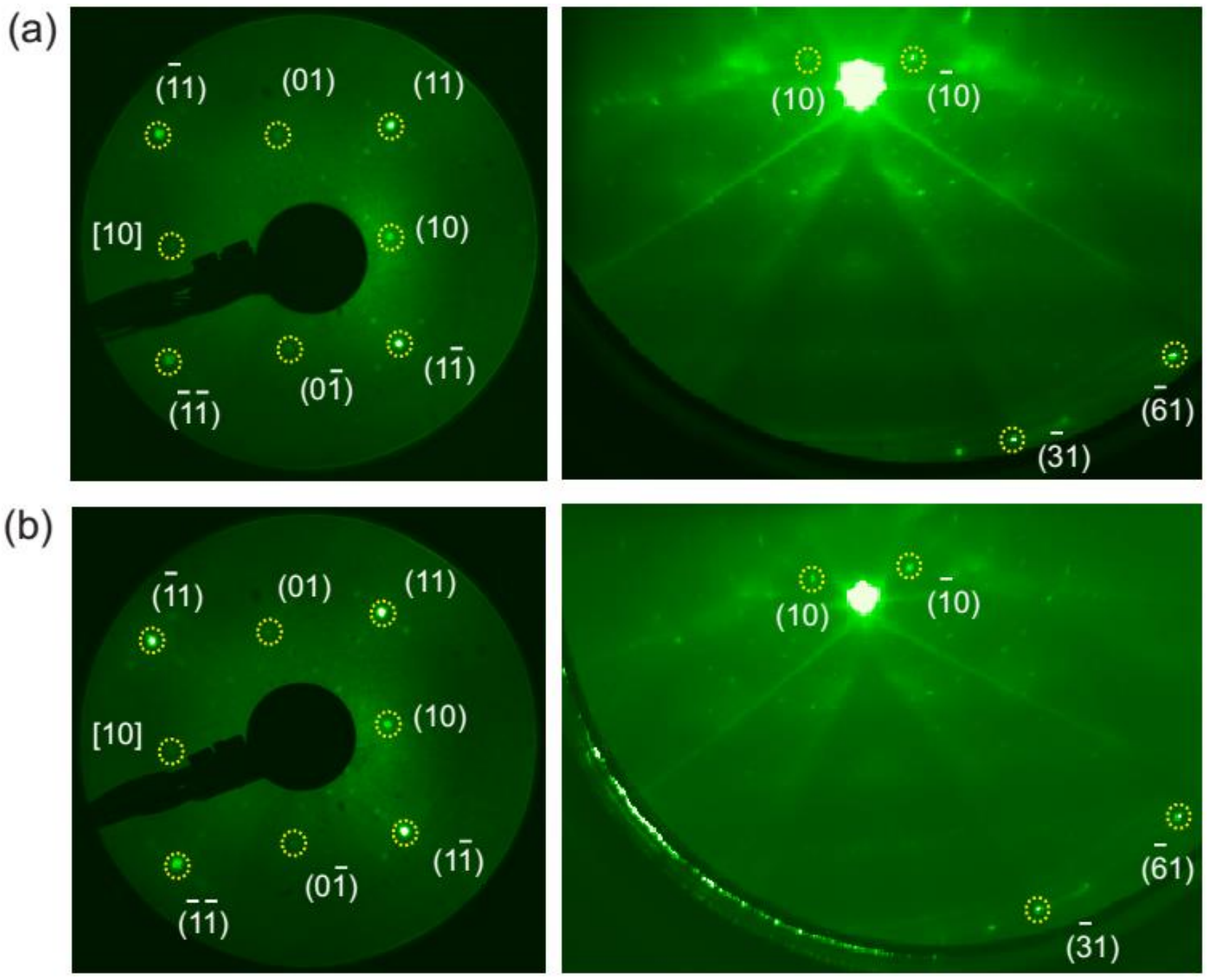

Figure S1. LEED patterns at 100 eV beam energy and RHEED patterns along the [010] direction for air-annealed (a) and $O_2$-anealed (b) ‘as-loaded’ $TiO_2$(001) substrates. There was no further surface treatment after loading samples to the UHV system. Integer order spots

are indexed in yellow circles. Faint fractional order spots of the $c(7\sqrt{2}\times5\sqrt{2})R45°$ reconstruction are already visible after both furnace annealing treatments.

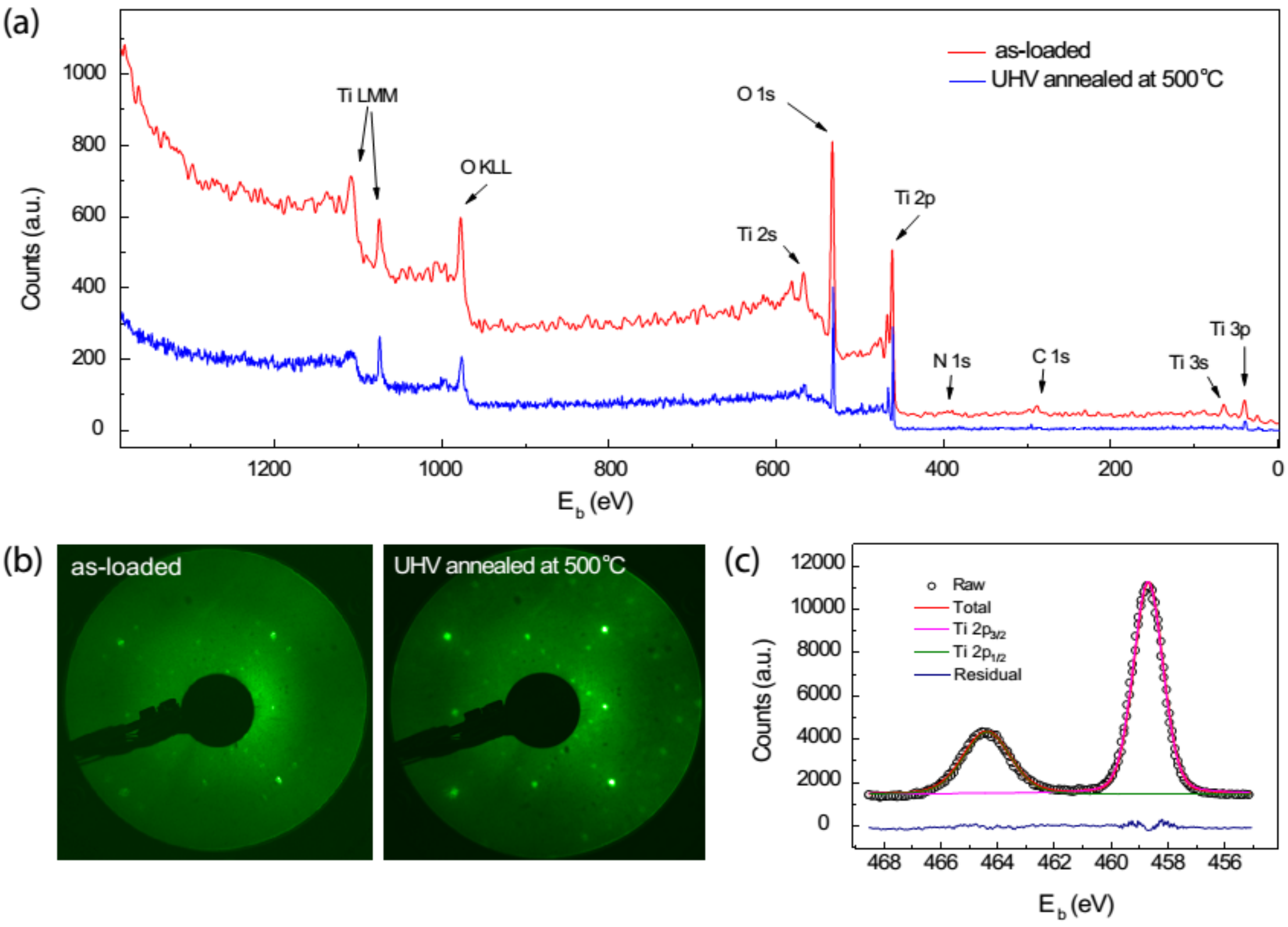

Figure S2. Survey X-ray Photoelectron spectra (a) and LEED patterns at 100 eV (b) of an as-loaded and subsequently UHV annealed $TiO_2$(001) substrate. (c) Ti 2*p* core level spectrum of the as-loaded sample. The survey spectrum does not show sufficient nitrogen, carbon or other species to conclude that N or C species are incorporated in the unit cell of the observed reconstruction (compare to Figs. 1(b) in Refs. 4 and 34. The Ti 2*p* core level spectrum (c) can be well fitted using a single core level component, indicating there is no reduced $Ti^{3+}$ on the surface and the surface is fully oxidized to $Ti^{4+}$. These results verify that the $c(7\sqrt{2}\times5\sqrt{2})R45°$ reconstruction is not stabilized by impurities introduced during the furnace anneal.

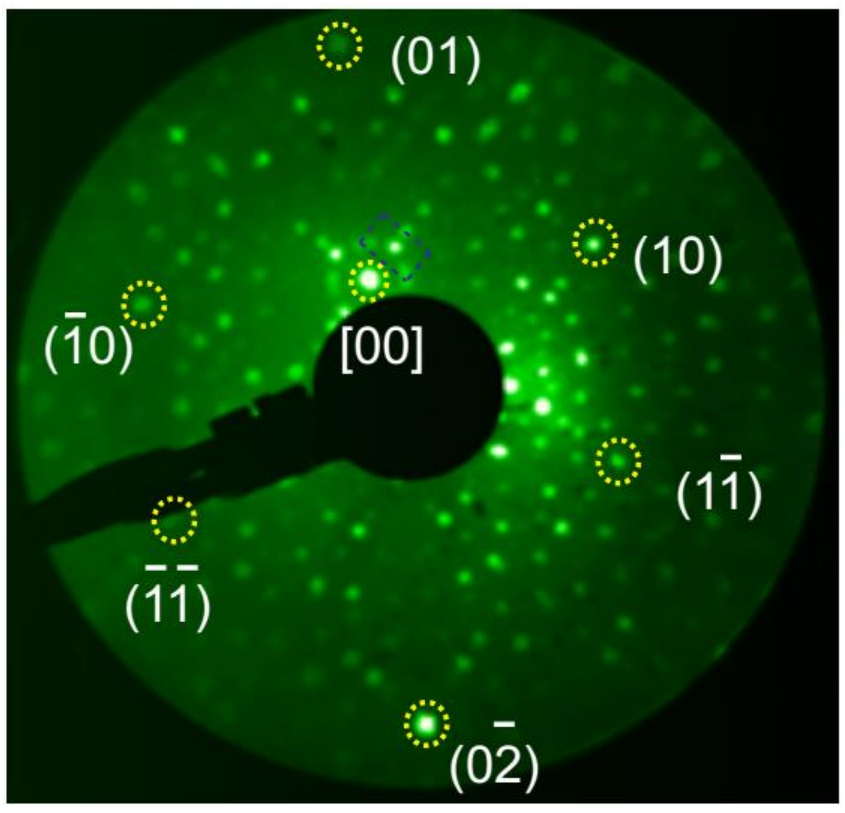

Figure S3. LEED pattern at 82 eV beam energy after oxygen plasma exposure at 500 °C for 40 minutes. The incident electron beam is slightly off-normal so as to reveal the fractional

order spots near the (0,0) position. A unit cell of the $c(7\sqrt{2}\times5\sqrt{2})R45°$ superstructure is indicated.

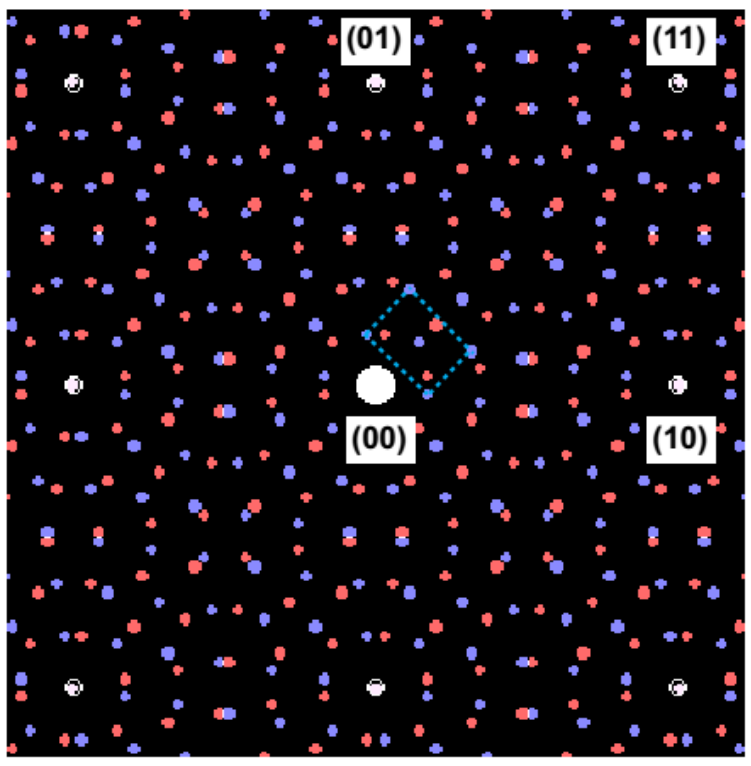

Figure S4. Simulated reciprocal space lattice of the $c(7\sqrt{2}\times5\sqrt{2})R45°$ reconstruction based on the square (p4/mmm) 1×1 substrate lattice. Blue and red dots originate from the two orthogonal domains, as dictated by the substrate symmetry. A unit cell of the $c(7\sqrt{2}\times5\sqrt{2})R45°$ superstructure is indicated by a blue dashed rectangle.

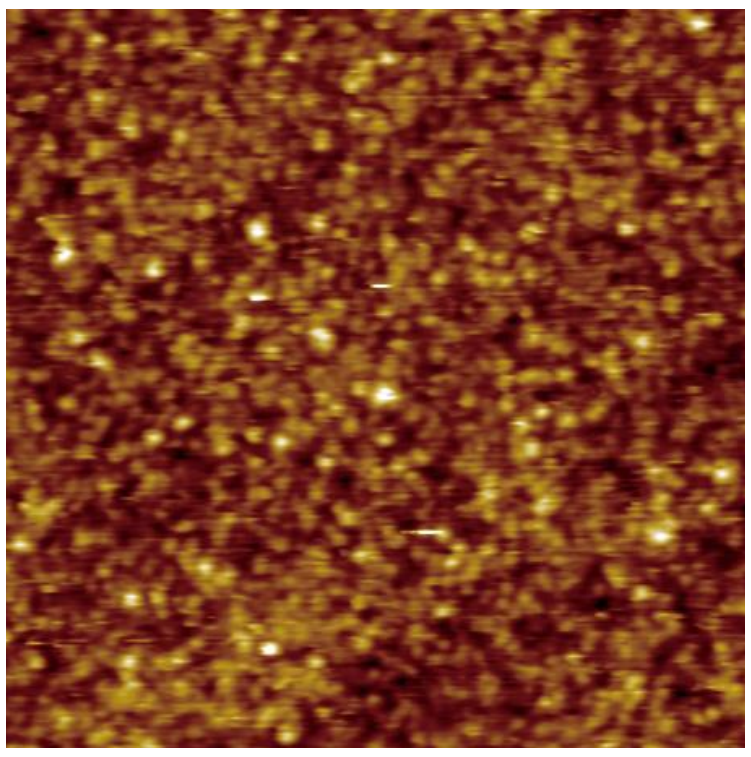

Figure S5. STM image (+5.0 V, 5.0 pA, 50 nm × 50 nm) of the as-loaded $TiO_2$(001) surface after degassing.

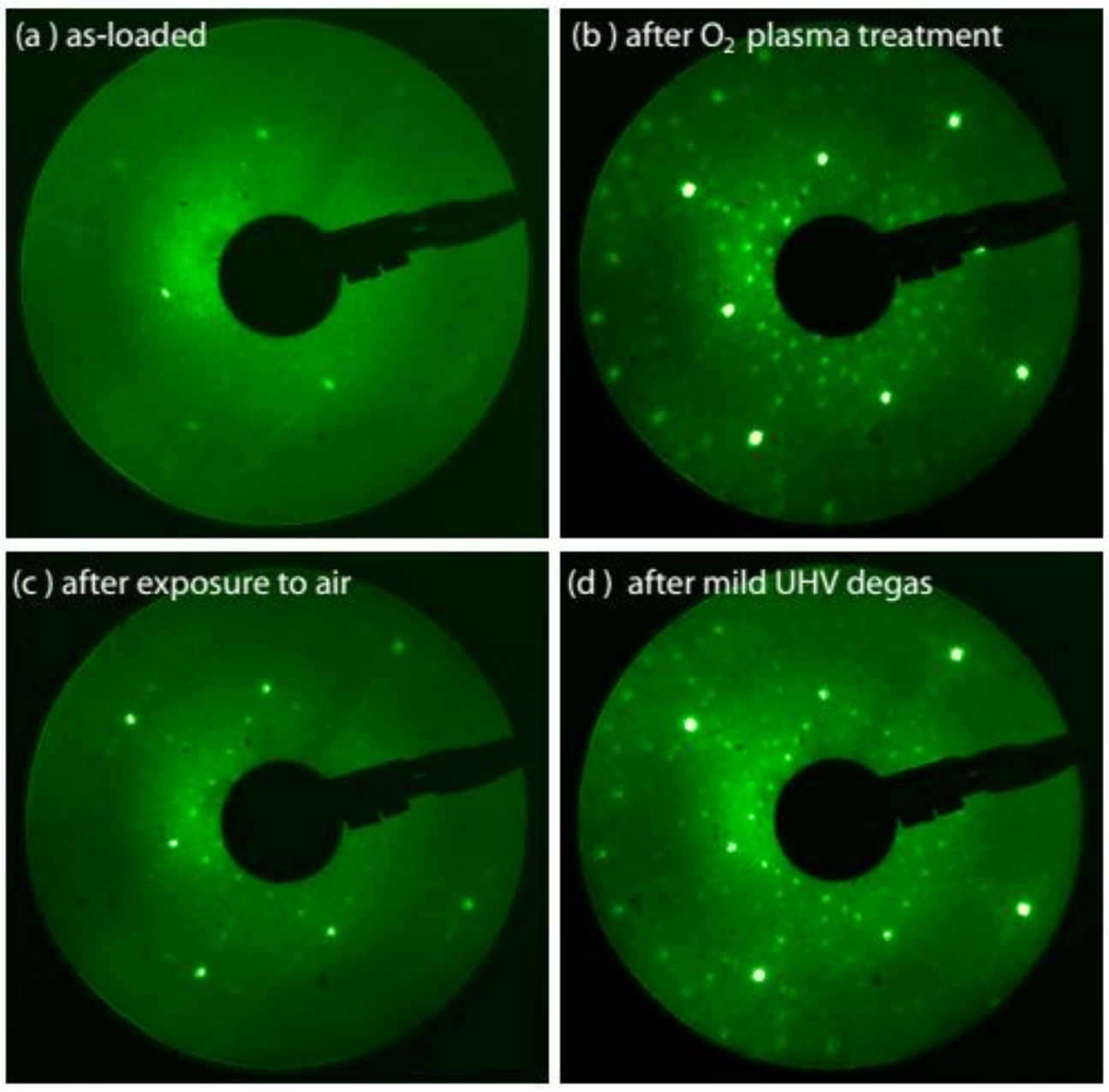

Figure S6. LEED patterns of $TiO_2(001)$ substrates for an as-loaded sample (a); after annealing in an $O_2$ plasma at 500 °C for 40 min (b); after subsequent exposure to air and reloading into the vacuum system (c); and finally after a mild degassing in UHV (d). These data show that the $c(7\sqrt{2}\times5\sqrt{2})R45°$ reconstruction persists even after exposure to air.